\documentstyle[preprint,epsfig,aps,12pt]{revtex}
\tighten
\begin{document}
\bibliographystyle{revtex}
\title{Microscopic optical potentials for nucleon-nucleus scattering}
\author{S. Karataglidis and M. B. Chadwick}
\address{Theoretical Division, Los Alamos National Laboratory, Los
Alamos, NM, 87545}
\date{\today}
\preprint{LA-UR-01-1849}
\maketitle
\begin{abstract}
Microscopic optical potentials for nucleon-nucleus ($NA$) scattering
obtained from the folding of the effective $g$ matrices,
solutions of the Bruckner-Bethe-Goldstone equation, with the densities
of the target, are applied to the case of neutron-nucleus
scattering. Given that the optical potentials are specified in all
two-body angular momentum, spin and isospin channels available to the
$NA$ scattering, the only difference in this model description of
proton and neutron scattering observables for a given nucleus is in
the inclusion of the Coulomb interaction. {\em A priori} microscopic
optical model predictions for neutron and proton elastic scattering
are compared with results from a phenomenological optical model and
with data. New measurements are recommended to reduce discrepancies in
the existing database, and to differentiate between different
theoretical predictions.
\end{abstract}

\section{Introduction}
All of the development of the folding models of the microscopic
nucleon-nucleus ($NA$) optical potentials for medium energy scattering
to date have concentrated on the description of proton-nucleus ($pA$)
scattering observables, with particular emphasis on studying the
effective interaction and its components (see Ref. \cite{Review} and
references cited therein). Much success in the description of
differential cross sections and spin observables has been achieved by
using a folding model in the coordinate space representation when
coupled with reasonable models of nuclear structure \cite{Do98}. More
recently, this model has also been applied to the predictions of
integral observables, with success in describing data from both proton
and neutron scattering \cite{De01}.

A natural extension of the application of this formalism would be to
describe the observables for neutron-nucleus ($nA$)
scattering. Comparison of $nA$ and $pA$ data for a given target and at
a given energy would elicit details specifically of the nuclear part
of the microscopic optical potential. Such a comparison would also
highlight the role of the Coulomb interaction in $NA$ scattering. It
is only the inclusion (or exclusion) of the Coulomb interaction being
the only difference in proton and neutron scattering that results in
the prediction that the analyzing powers for neutron and proton
elastic scattering from $^{208}$Pb at 100~MeV would be completely out
of phase \cite{Ko90}. This phenomenological result has yet to be
experimentally verified so it would be instructive to note if a
completely microscopic model for the scattering would lead to the same
conclusion. Such was done for 100~MeV scattering by Karataglidis and
Madland \cite{Ka01}, whose results supported the phenomenological
result.

Reaction and scattering information can be obtained directly from the
optical model. However, while numerous measurements have been made for
proton reaction cross sections and elastic scattering distributions,
very few measurements have been made for neutrons above 20~MeV, owing
to experimental difficulties in producing suitable monoenergetic
neutron beams. Thus, phenomenological proton optical potentials can be
obtained from fitting the available proton scattering data, but this
approach cannot be directly used to obtain phenomenological neutron
potentials, though procedures exist by which the neutron potential may
be obtained from that of the proton using a Lane model. Therefore, we
look to the applicability of the microscopic $NA$ optical potentials
to $nA$ scattering.

A microscopic theory of $NA$ scattering must necessarily start with an
appropriate form of the underlying nucleon-nucleon ($NN$) interaction
from which the $g$ matrices for nucleons scattering from infinite
nuclear matter are obtained as solutions of the
Bruckner-Bethe-Goldstone (BBG) equation. A local density approximation
then maps the $g$ matrix onto an effective $NN$ interaction in medium
for the target in question. That effective interaction is folded with
density of the target to obtain the microscopic, nonlocal, optical
potential, cast in terms of central, tensor and spin-orbit terms. For
the calculations presented herein, we use the Bonn B \cite{Ma87} $NN$
interaction as the starting interaction, with densities obtained
primarily from the shell model, specifying both the density dependence
of the effective $NA$ interaction and the density of the target.

The examples considered herein are 65~MeV proton and neutron
scattering from $^{12}$C, $^{28}$Si, $^{40}$Ca, $^{56}$Fe, $^{90}$Zr,
and $^{208}$Pb. These choices are predicated on the availability of
both proton and neutron scattering data. Comparisons of the results of
the microscopic calculations are also made with results obtained using
a global phenomenological optical potential applicable for the mass
range encompassed in our investigation.

\section{Optical potentials}
The complete details of the calculation of the microscopic optical
potential can be found in a recent review article \cite{Review},
including details of the program DWBA98 \cite{Ra99}, with which we
calculate the optical potential and all observables. Herein, we
present a summary showing the relevant details to allow for a
comparison of $pA$ and $nA$ scattering.

With $\vec{r}$ and $\vec{r}'$ denoting the relative coordinates between
a colliding pair of particles, the Schr\"odinger equation describing
their scattering by a local Coulomb, $V_C(r)$, and a nonlocal hadronic
(optical) potential is of the form
\begin{equation}
\left[ \frac{\hbar^2}{2\mu}\nabla^2 - V_C(r) + E \right] \Psi(\vec{r})
= \int U(\vec{r}, \vec{r}') \Psi( \vec{r}' ) d\vec{r}' \;,
\label{cat}
\end{equation}
where $\Psi(\vec{r})$ is the scattering solution and $U(\vec{r},
\vec{r}')$ is the optical potential. The optical potential is obtained
by the folding of the relevant nuclear structure information with the
effective $g$ matrices, as specified in $ST$-channel form and in
coordinate space. They are obtained from the set of infinite matter
nuclear $g$ matrices \cite{Ha70,Do91} obtained by solution of the BBG
equation,
\begin{eqnarray}
g^{(JST)}_{LL'}(p', p; k, K, k_F) & = & V^{(JST)}_{LL'}(p,p') \nonumber \\
& + & \frac{2}{\pi} \sum_l \int^{\infty}_0 V^{(JST)}_{Ll}(p',q) \left[
{\cal H} \right]  g^{(JST)}_{lL'}(q,p; k, K, k_F) q^2 dq\;,
\label{gmat}
\end{eqnarray}
where
\begin{equation}
{\cal H}( q, k, K, k_F ) = \frac{ \bar{Q}(q,K,k_F) }{ \bar{E}( k, K,
k_F ) - \bar{E}( q, K, k_F ) + i\varepsilon }
\end{equation}
in which $\bar{Q}(q,K,k_F)$ is an angle-averaged Pauli operator with
an average center-of-mass momentum $K$ \cite{Ha70,Do91}. The energies
in the propagator of the BBG equation include auxiliary potentials $U$
\cite{Ha70,Do91} (first order mass operator) and are defined by
\begin{equation}
\bar{E}( q, K, k_F ) = (q^2 + K^2) + \left( \frac{m}{\hbar^2} \right)
\left\{ U\left( \left| \vec{q} + \vec{K} \right| \right) + U\left(
\left| \vec{q} - \vec{K} \right| \right) \right\} \;.
\end{equation}

The nuclear structure information is specified in terms of the
one body density matrix elements (OBDME) (with $\alpha$ specifying the
set $n, l, j$ and $\tilde{a}_{\alpha m} = (-1)^{j-m}a_{\alpha -m}$)
\begin{equation}
S_{\alpha \alpha' I} = \left\langle J_f \left\| \left[
a^{\dag}_{\alpha'} \times \tilde{a}_{\alpha} \right]^I \right\| J_i
\right\rangle \rightarrow \left\langle J \left\| \left[
a^{\dag}_{\alpha'} \times \tilde{a}_{\alpha} \right]^I \right\| J
\right\rangle,
\end{equation}
for the case of elastic scattering from a target of spin $J$. With the
diagonal OBDME specified in the occupation number representation, for
$I=0$,
\begin{equation}
S_{\alpha \alpha 0} = \sqrt{2j+1} \sigma_{\alpha \alpha}
\end{equation}
where $\sigma_{\alpha \alpha}$ is the fractional shell occupancy with
a filled shell signifying $\sigma_{\alpha \alpha} = 1$, the optical
potential given by the folding process, takes the form
\begin{eqnarray}
U( \vec{r}_1, \vec{r}_2; E ) & = & \sum_{\alpha m \alpha' m'} (2j+1)
\sigma_{\alpha \alpha'} \nonumber \\
& \times & \left[ \delta( \vec{r}_1 - \vec{r}_2 ) \int
\varphi^{\ast}_{\alpha' m'}(\vec{s}) U^D\left( R_{1s}, E \right)
\varphi_{\alpha m}( \vec{s} ) d\vec{s} + \varphi^{\ast}_{\alpha'
m'}( \vec{r}_1 )U^{Ex}\left( R_{12}, E \right) \varphi_{\alpha m}(
\vec{r}_2 ) \right],
\end{eqnarray}
where $R_{12} = \left| \vec{r}_1 - \vec{r}_2 \right|$ and $R_{1s} =
\left| \vec{r}_1 - \vec{s} \right|$, $\phi_{\alpha m}$ are the single
particle (SP) wave functions specifying the nucleons, and $U^D$ and
$U^{Ex}$ are appropriate combinations of the multipoles of the
effective interactions for the direct and exchange contributions to
the optical potential respectively. The exchange term arises from the
antisymmetry of the projectile and the struck (bound) nucleon within
the nucleus.

The exchange terms are the major component of nonlocality in the $NA$
optical potential. Indeed, neglect of those terms specifically in
folding models leads to severe problems in the description of the
observables \cite{De00}, with the differential cross sections being
overpredicted in some cases by up to several orders of
magnitude. Another source of nonlocality is the $NN$ interaction
itself. In the calculation of $g$ matrices, that
nonlocality also contributes to that of the optical
potential and manifests itself partly in the energy and density
dependences in those $g$ matrices. That it contributes to the overall
nonlocal nature of the optical potential is also evident from the
off-shell part of the $g$ matrices. The extrapolations of the $g$
matrices off-shell, relative to their on-shell values, are similar to
those for the free $NN$ $t$ matrices \cite{Review}, and so all aspects
of the $NN$ interaction are carried through in the solution of the BBG
equations. We define such a model as $g$-folding.

It is clear from Eq.~(\ref{gmat}) that the appropriate form of the
optical potential for $pA$ and $nA$ scattering calculations may be
obtained from the common $g$ matrix, with the projectile isospin
allowing the selection of the correct components through the appropriate
two-body isospin channels. The only addition in the case of $pA$ scattering
is the inclusion of the Coulomb interaction in Eq.~(\ref{cat}).

The microscopic neutron and proton scattering predictions are compared
with predictions from a phenomenological optical potential developed
by Madland \cite{Ma97}. This potential is global in projectile energy,
isospin, and target $(Z,A)$, uses Woods-Saxon form factors, and was
developed for use in a relativistic Schr\"odinger (relativistic
kinematics) approach. The potential extends the earlier Schwandt work
\cite{Sc82} by increasing the target mass range from $A=24-208$ to
$A=12-208$, increasing the energy range from $E=80-180$~MeV to
$E=50-400$~MeV, and transforming the Schwandt proton potential to a
projectile-isospin dependent potential for neutrons and protons, using
a relativistic Lane model naturally incorporating Coulomb effects.
Experimental data for integrated observables (proton reaction, and
neutron total cross sections) and proton elastic scattering angular
distributions, were used to determine the phenomenological potential
parameters. The potential is described in detail in Ref.~\cite{Ma97}.

The observables are obtained once the optical potential has been
calculated, as is detailed in the recent review \cite{Review}. In
particular, the integral observables are obtained from the $S$ matrix,
or equivalently the phase shifts, $\delta_l(k)$:
\begin{equation}
S^{\pm}_l \equiv S^{\pm}_l(k) = e^{2i\delta^{\pm}_l(k)} =
\eta^{\pm}_l(k)e^{2i\Re\left[ \delta^{\pm}_l(k) \right] }
\end{equation}
where
\begin{equation}
\eta^{\pm}_l \equiv \eta^{\pm}_l(k) = \left| S^{\pm}_l(k) \right| =
e^{-2\Im\left[ \delta^{\pm}_l(k) \right] } \;.
\end{equation}
In terms of these and with $E \propto k^2$, the elastic and total reaction
(absorption) cross sections are given by
\begin{eqnarray}
\sigma_{\mbox{el}}(E) & = & \frac{\pi}{k^2} \sum^{\infty}_{l=0}
\left\{ (l+1) \left| S^+_l -1 \right|^2 + l \left| S^-_l -1
\right|^2 \right\} \;, \nonumber \\
\sigma_R(E) & = & \frac{\pi}{k^2} \sum^{\infty}_{l=0} \left\{ (l+1)
\left[ 1 - \left( \eta^+_l \right)^2 \right] + l \left[ 1 - \left(
\eta^-_l \right)^2 \right] \right\} \;,
\end{eqnarray}
respectively. The total cross section, $\sigma_{\mbox{\tiny TOT}}(E)$,
is the sum of these two cross sections.

\section{Results}
Select results are presented for elastic 65~MeV $NA$ scattering from
$^{12}$C, $^{28}$Si, $^{40}$Ca, $^{56}$Fe, $^{90}$Zr, and
$^{208}$Pb. We present the differential cross sections and analysing
powers for all cases, as well as consider the integral neutron
scattering data: the total, elastic, and total reaction cross
sections. The shell model calculations, where applicable, were
performed using the code OXBASH \cite{Ox86}. The starting $NN$
interaction for the microscopic calculations was the Bonn B
interaction \cite{Ma87}.

The SP wave functions for the microscopic calculations were chosen to
be harmonic oscillators, with the oscillator parameter chosen to
reproduce the root-mean-square (r.m.s.) radius of each nucleus within
the given shell model space. The exception to that was $^{208}$Pb, for
which we used a Skyrme-Hartree-Fock (SHF) calculation \cite{Br00}. In
that calculation, the neutron wave functions were chosen such that the
difference in the neutron and proton r.m.s. radii were $0.16 \pm
0.02$~fm, the Friedman-Pandharipande neutron equation of state serving
as the constraint. The choice of SHF wave functions reproduce both the
proton and predicted neutron r.m.s. radius. For the other nuclei,
Table~\ref{radii} lists the oscillator parameter used for each nucleus
along with the shell model used, and predicted r.m.s. radius. In all
cases, the models used together with the choice of oscillator
parameters reproduce the r.m.s. radii quite well. For all nuclei,
these models and SP wave functions were used in the microscopic
calculations of the scattering presented below.

\begin{figure}
\centering\epsfig{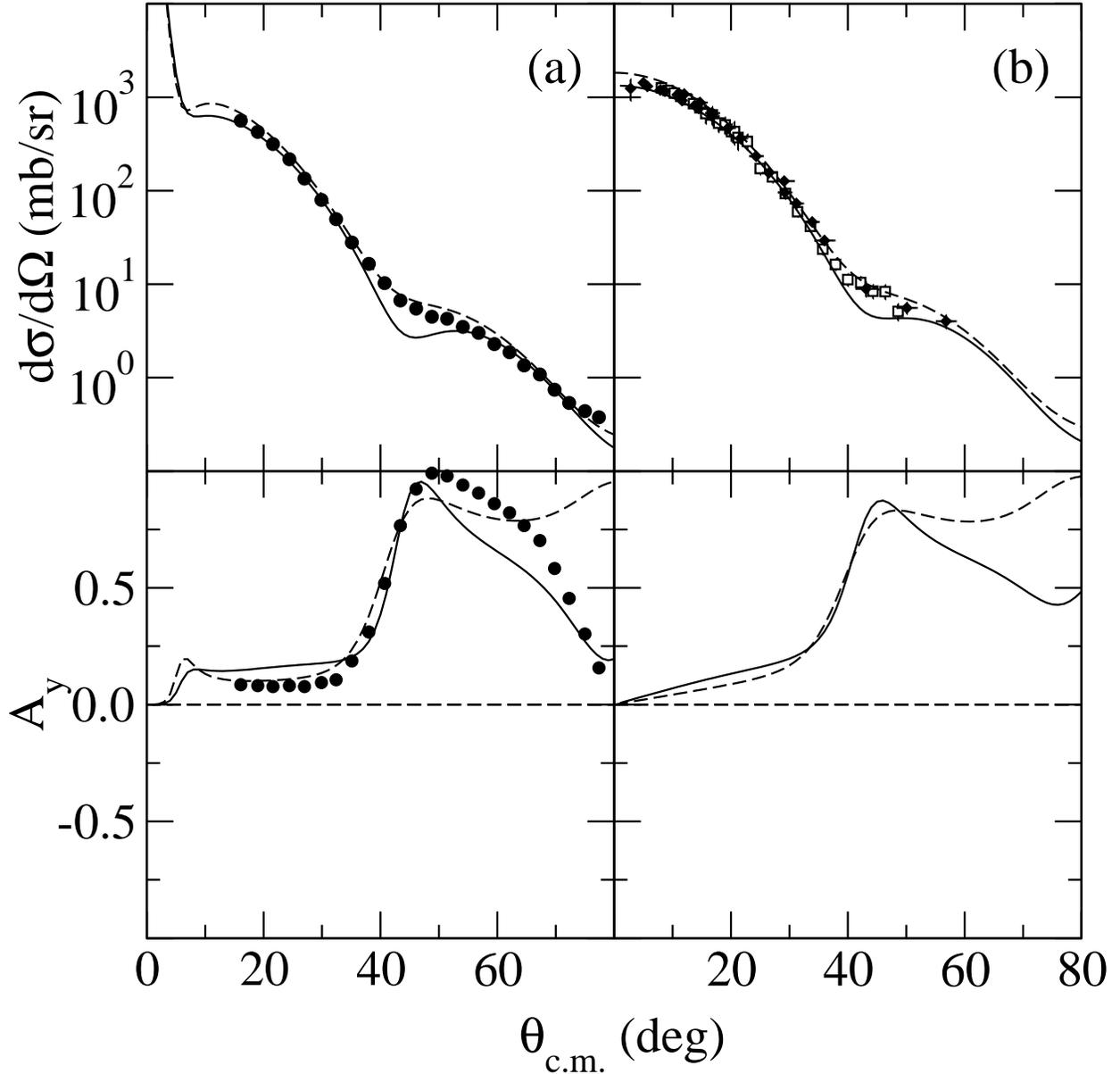}
\caption[]{Differential cross sections and analysing powers for the
scattering of 65~MeV protons (a) and neutrons (b) from $^{12}$C. The
proton scattering data of Kato {\em et al.} \cite{Ka85} (circles), and
the neutron scattering data of Hjort {\em et al.}  (squares)
\cite{Hj94} and Ibaraki {\em et al.} (diamonds) \cite{Ib00} are
compared to the results of the calculations made using the microscopic
(solid line) and phenomenological (dashed line) optical potentials.}
\label{carbon}
\end{figure}

The results of the microscopic calculations of the scattering from
$^{12}$C, as well as those obtained from the phenomenological model,
are compared to the available data in Fig.~\ref{carbon}.  Therein, the
proton scattering data of Kato {\em et al.} \cite{Ka85} and the
neutron scattering data of Hjort {\em et al.} \cite{Hj94} (squares)
and Ibaraki {\em et al.} \cite{Ib00} (diamonds) are compared to the
results of the calculations made using the microscopic and
phenomenological optical potentials. While the microscopic potential
results tend to underestimate the region of the minimum ($\sim
40^{\circ}$), for both proton and neutron scattering from $^{12}$C,
the model reproduces well the overall shape and magnitude. The model
also reproduces the analyzing powers well, and predicts the negative
slope above $60^{\circ}$, which is not present in the result from the
phenomenological calculation.  For neutron scattering, the microscopic
model gives much better agreement with the data for the forward angle
cross section, and so one expects that the total cross section would
be better reproduced by this model. That total elastic cross section,
along with the total reaction and total neutron scattering cross
sections for all the nuclei, are given in Table~\ref{cross}. As
expected, the total cross section for $n$-$^{12}$C scattering at
65~MeV is much better reproduced by the microscopic result. Also,
there are differences between the predicted total elastic and total
reaction cross sections. The microscopic model predicts a smaller
elastic cross section and larger reaction cross section than the
phenomenological model, suggesting the microscopic model calculates a
weaker real and stronger imaginary part of the potential. Note that
the total neutron scattering cross section from $^{12}$C predicted by
the microscopic model at intermediate energies is in good agreement
with the data \cite{De01}.

\begin{figure}
\centering\epsfig{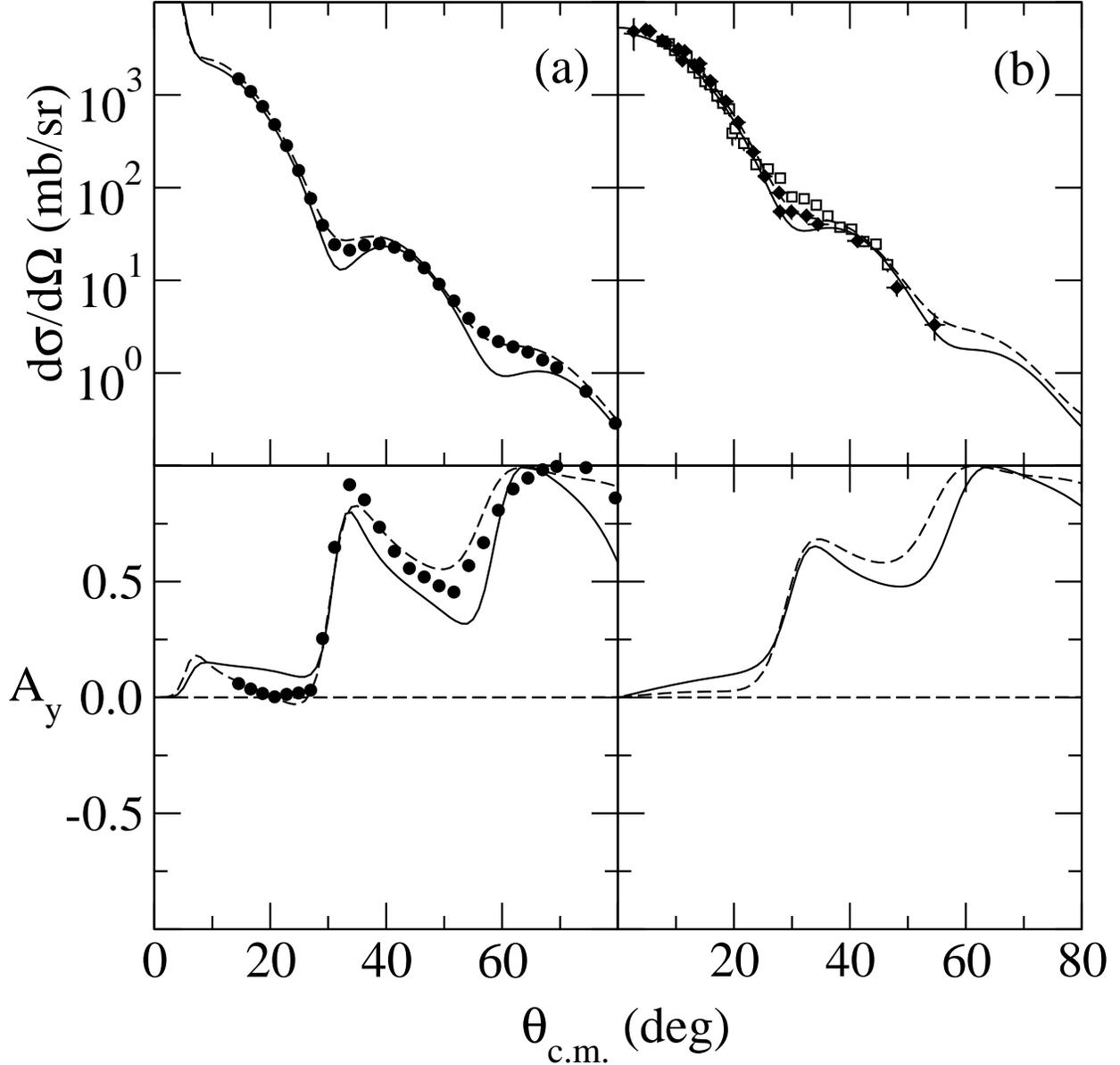}
\caption[]{As for Fig.~\ref{carbon}, but for $^{28}$Si. The proton
scattering data [circles, (a)] are those of Sakaguchi {\em et al.}
\cite{Sa82}, while the neutron scattering data (b) are from Hjort {\em et al.}
\cite{Hj94} (squares) and Ibaraki {\em et al.} \cite{Ib00}
(diamonds).}
\label{silicon}
\end{figure}

We compare the results of our calculations for the 65~MeV scattering
of nucleons from $^{28}$Si to the data in Fig.~\ref{silicon}. The
proton scattering data are those of Sakaguchi {\em et al.}
\cite{Sa82}, while the neutron data are those of Hjort {\em et al.}
\cite{Hj94} and Ibaraki {\em et al.} \cite{Ib00}. The proton
scattering data are described well by both models, the
phenomenological model giving a better representation of both the
cross section and analyzing power.  There is a discrepancy between the
two sets of neutron scattering data around the first minimum, at
$25^{\circ}$. Both model results favor the Ibaraki set, giving
excellent reproduction of those data, including in the region of the
discrepancy. The models give similar results for the neutron analyzing
power. In Table~\ref{cross}, we find similar behavior in the optical
potentials for $^{28}$Si as for $^{12}$C, as evidenced by the relative
elastic and total reaction cross sections. Both models predict the
total cross section to within 3\%.

\begin{figure}
\centering\epsfig{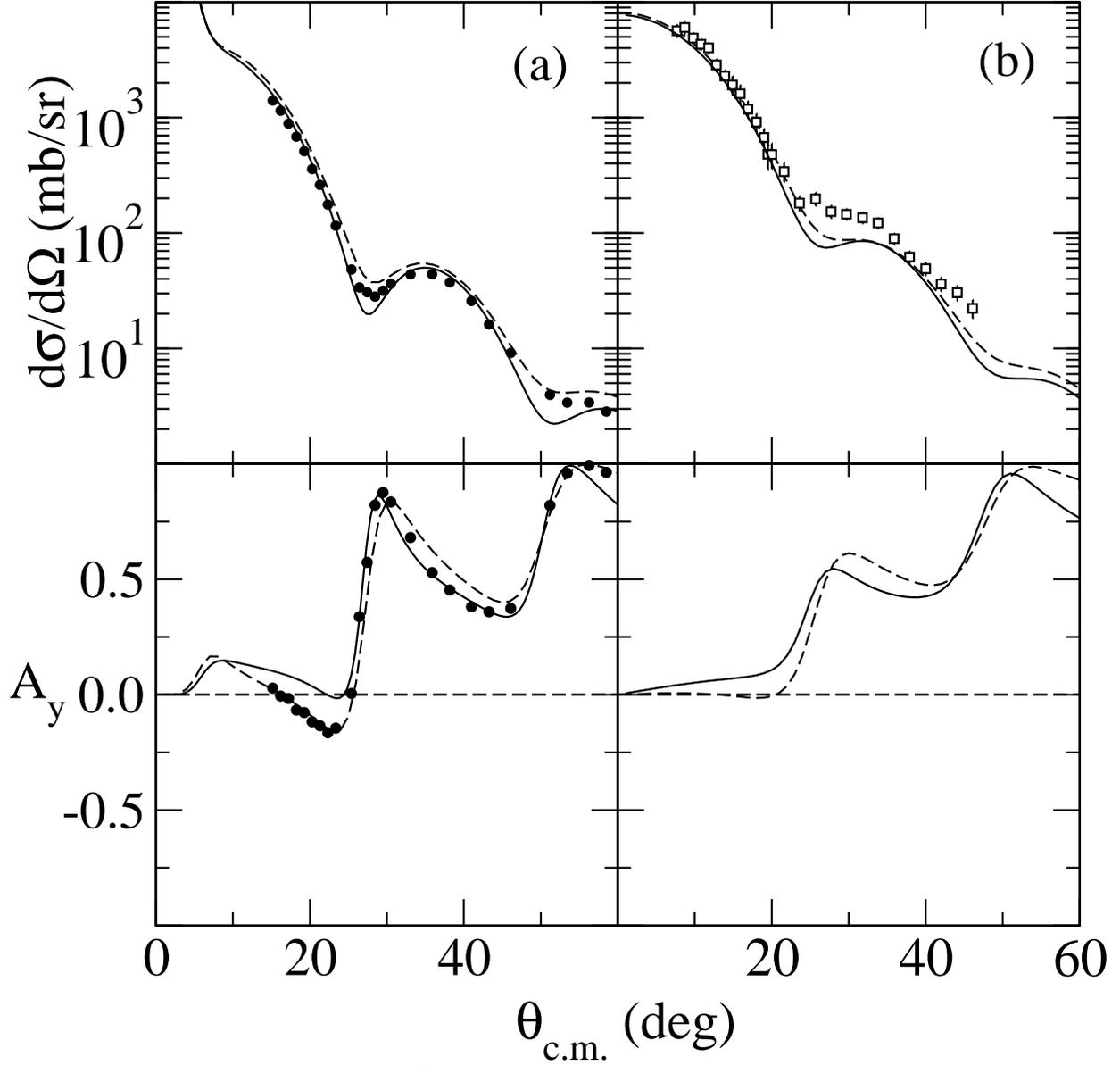}
\caption[]{As for Fig.~\ref{carbon}, but for $^{40}$Ca. The proton
scattering data [circles, (a)] are those of Sakaguchi {\em et al.}
\cite{Sa82}. The neutron data scattering (b) are those of Hjort {\em
et al.}  \cite{Hj94} (squares).}
\label{calcium}
\end{figure}

The differential cross sections and analyzing powers for the 65~MeV
scattering of nucleons from $^{40}$Ca are presented in
Fig.~\ref{calcium}. As with the previous results, both the microscopic
and phenomenological results agree with the data of Sakaguchi {\em et
al.} \cite{Sa82} quite well, with the phenomenology giving a better
representation of the observables. However, as with $^{28}$Si, both
calculations are unable to reproduce the neutron scattering
differential cross section data of Hjort {\em et al.} \cite{Hj94}, in
the region of the minimum, $\sim 23^{\circ}$. There are no other
available data for this case with which to compare.

\begin{figure}
\centering\epsfig{file=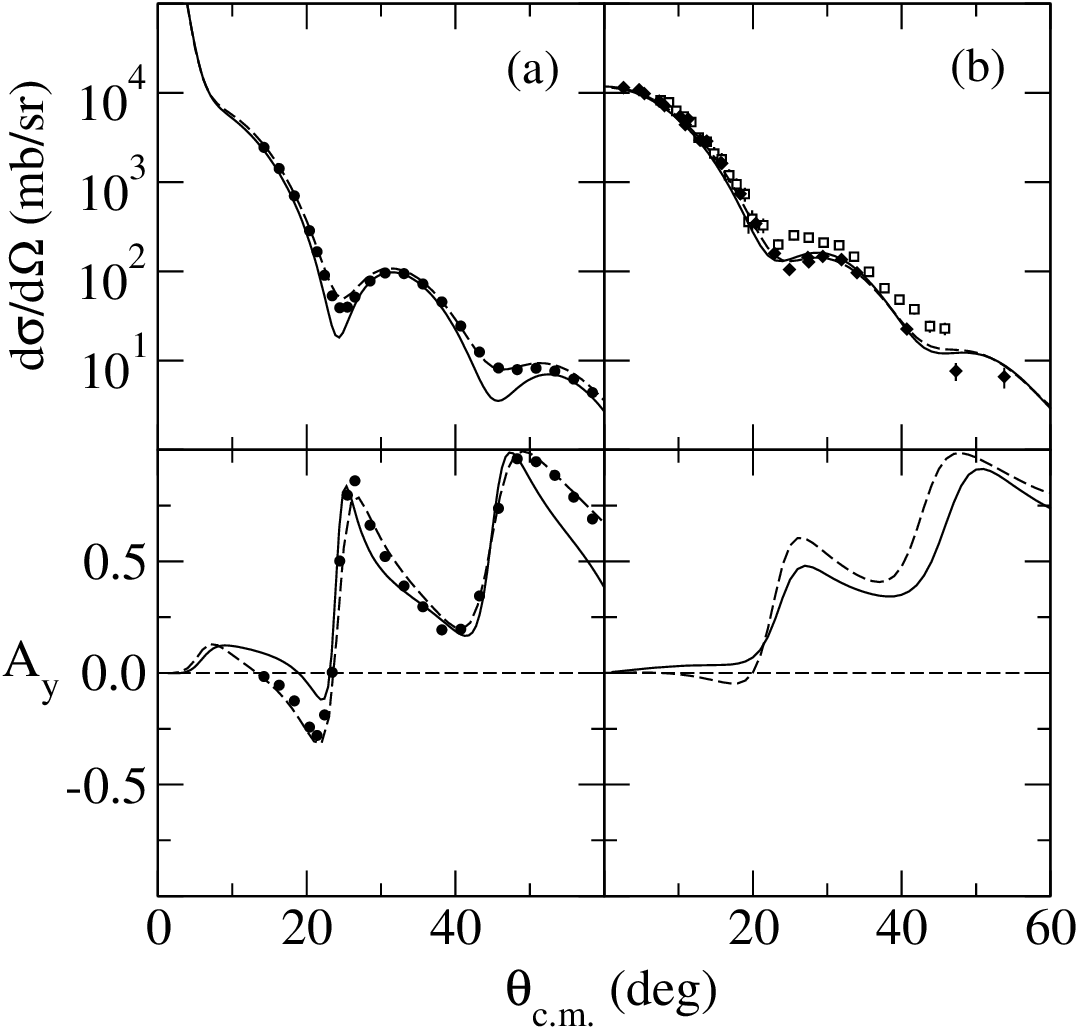,width=\linewidth,clip=}
\caption[]{As for Fig.~\ref{carbon}, but for $^{56}$Fe. The proton
scattering data [circles, (a)] are those of Sakaguchi {\em et al.}
\cite{Sa82}, while the neutron scattering data in (b) are those of
Hjort {\em et al.}  \cite{Hj94} (squares) and Ibaraki {\em et al.}
\cite{Ib00} (diamonds).}
\label{iron}
\end{figure}

In Fig.~\ref{iron}, we compare the results of both model calculations
for the scattering of 65~MeV nucleons from $^{56}$Fe to the proton
scattering data of Sakaguchi {\em et al.} \cite{Sa82}, and to the
neutron scattering data of Hjort {\em et al.} \cite{Hj94} and Ibaraki
{\em et al.} \cite{Ib00}. The agreement between the model calculations
and the data for proton scattering is quite good, and as before the
phenomenological result gives the better representation of the
data. As with $^{28}$Si, a discrepancy between the Hjort and Ibaraki
data sets exists in the region of the first minimum ($\sim
22^{\circ}$) of the neutron scattering differential cross section. The
model calculations both favor the Ibaraki set. There would appear to
be a normalization problem in the Hjort data in this region in
general, when one considers also the previous examples. This would
suggest the need for another measurement of these data in order to
resolve the discrepancy. In the case of the neutron integral
observables, the optical potentials for $^{56}$Fe reflect the same
behavior as for the lighter systems, although in this case the
predicted total cross sections are in agreement.

\begin{figure}
\centering\epsfig{file=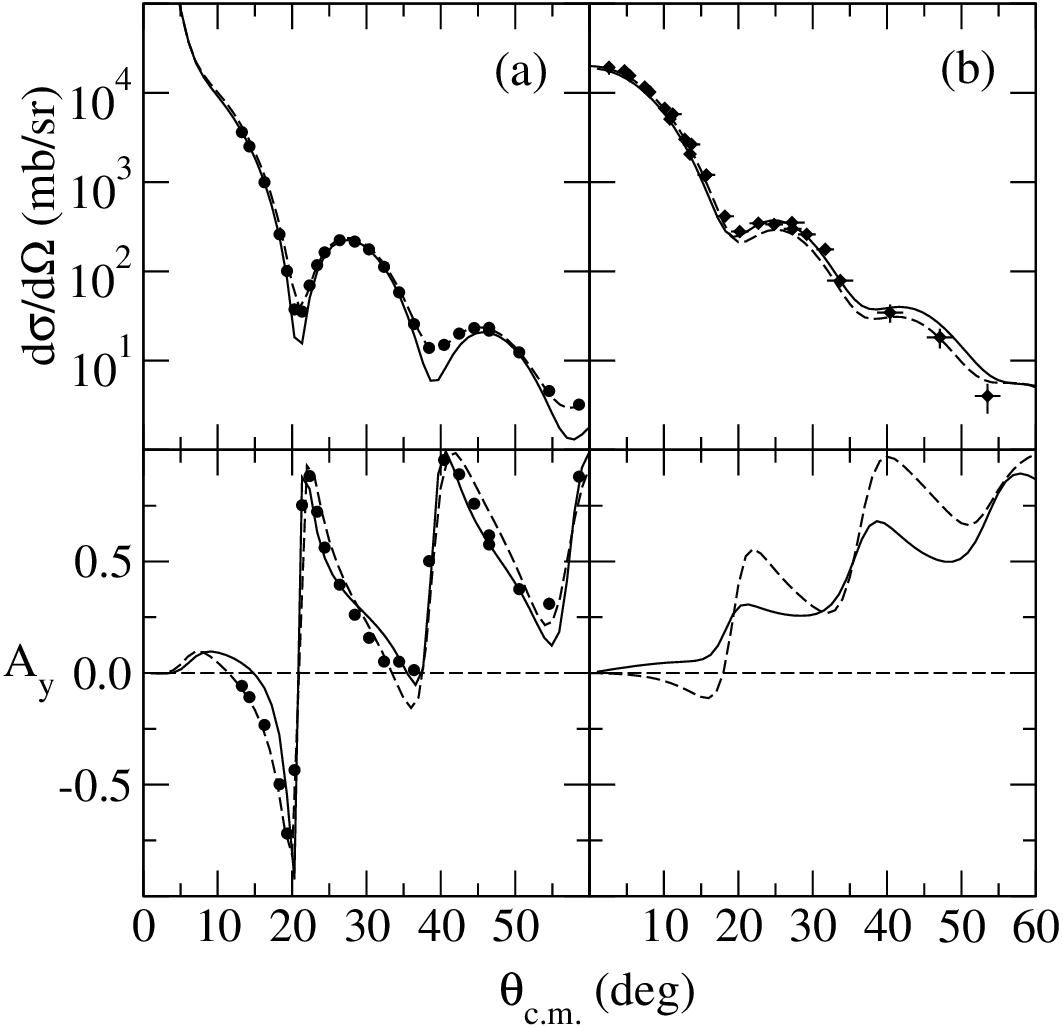,width=\linewidth,clip=}
\caption[]{As for Fig.~\ref{carbon}, but for $^{90}$Zr. The proton
scattering data [circles, (a)] are those of Sakaguchi {\em et al.}
\cite{Sa82}, while the neutron scattering data in (b) are those of
Ibaraki {\em et al.} \cite{Ib00} (diamonds).}
\label{zirconium}
\end{figure}

The results of both model calculations for the 65~MeV scattering of
nucleons from $^{90}$Zr are compared to the proton scattering data of
Sakaguchi {\em et al.} \cite{Sa82} and neutron scattering data of
Ibaraki {\em et al.} \cite{Ib00} in Fig.~\ref{zirconium}. There is
good agreement between the model results and the data. This is also
reflected in the total cross section, for which both models are in
agreement with the measured value to within 1\%. However, we would
encourage measurement of the neutron analyzing power in this
case. That analyzing power as predicted by both models is very much
different to the proton one, unlike those of the lighter nuclei. This
is an effect that was first observed by Kozack and Madland \cite{Ko90}
for 100~MeV nucleon scattering from $^{208}$Pb, and confirmed recently
by Karataglidis and Madland \cite{Ka01} using the microscopic
Schr\"odinger model.

\begin{figure}
\centering\epsfig{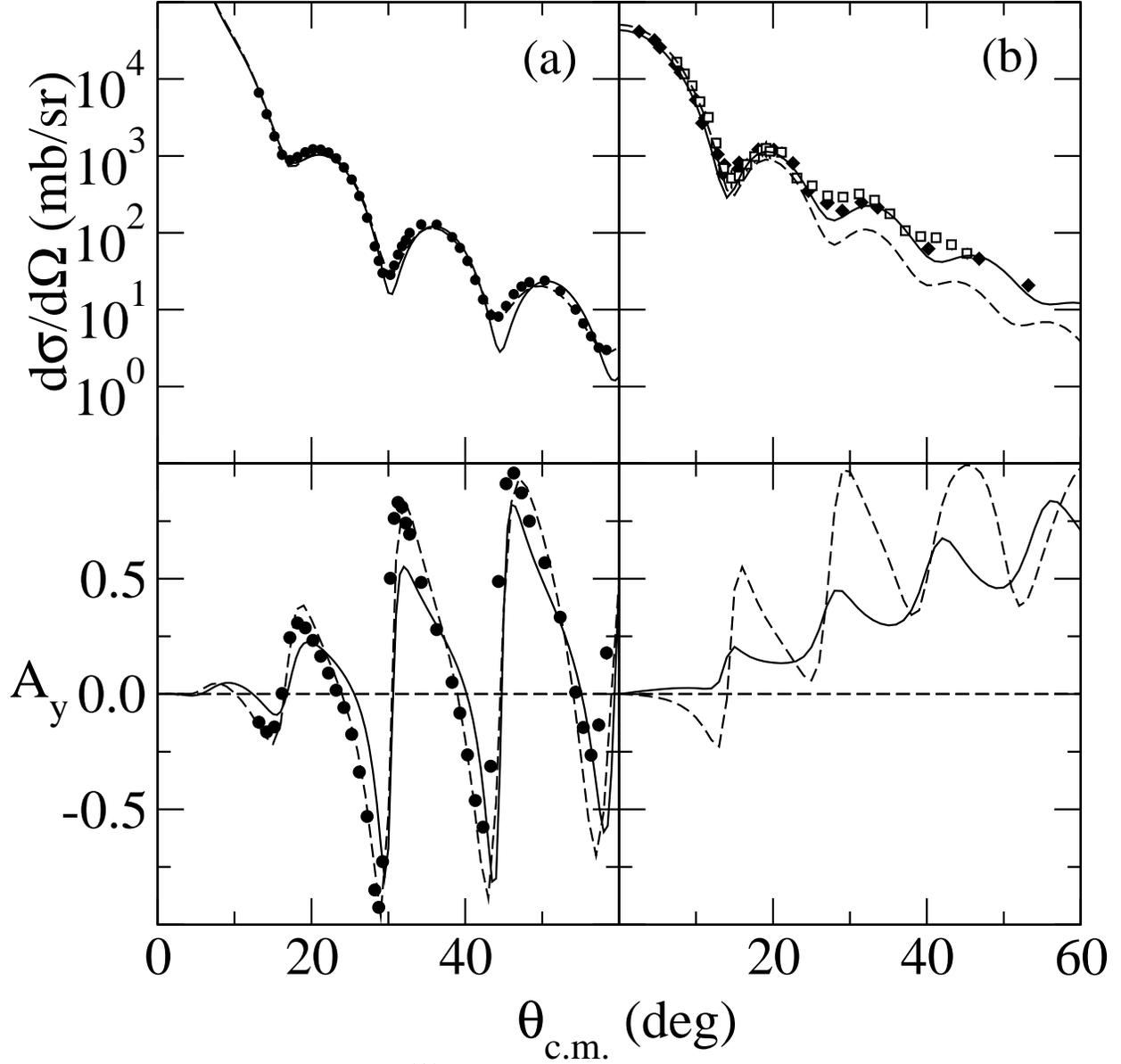}
\caption[]{As for Fig.~\ref{carbon}, but for $^{208}$Pb. The proton
scattering data [circles, (a)] are those of Sakaguchi {\em et al.}
\cite{Sa82}, while the neutron scattering data in (b) are those of
Hjort {\em et al.} \cite{Hj94} (squares) Ibaraki {\em et al.}
\cite{Ib00} (diamonds).}
\label{lead}
\end{figure}

\begin{figure}
\centering\epsfig{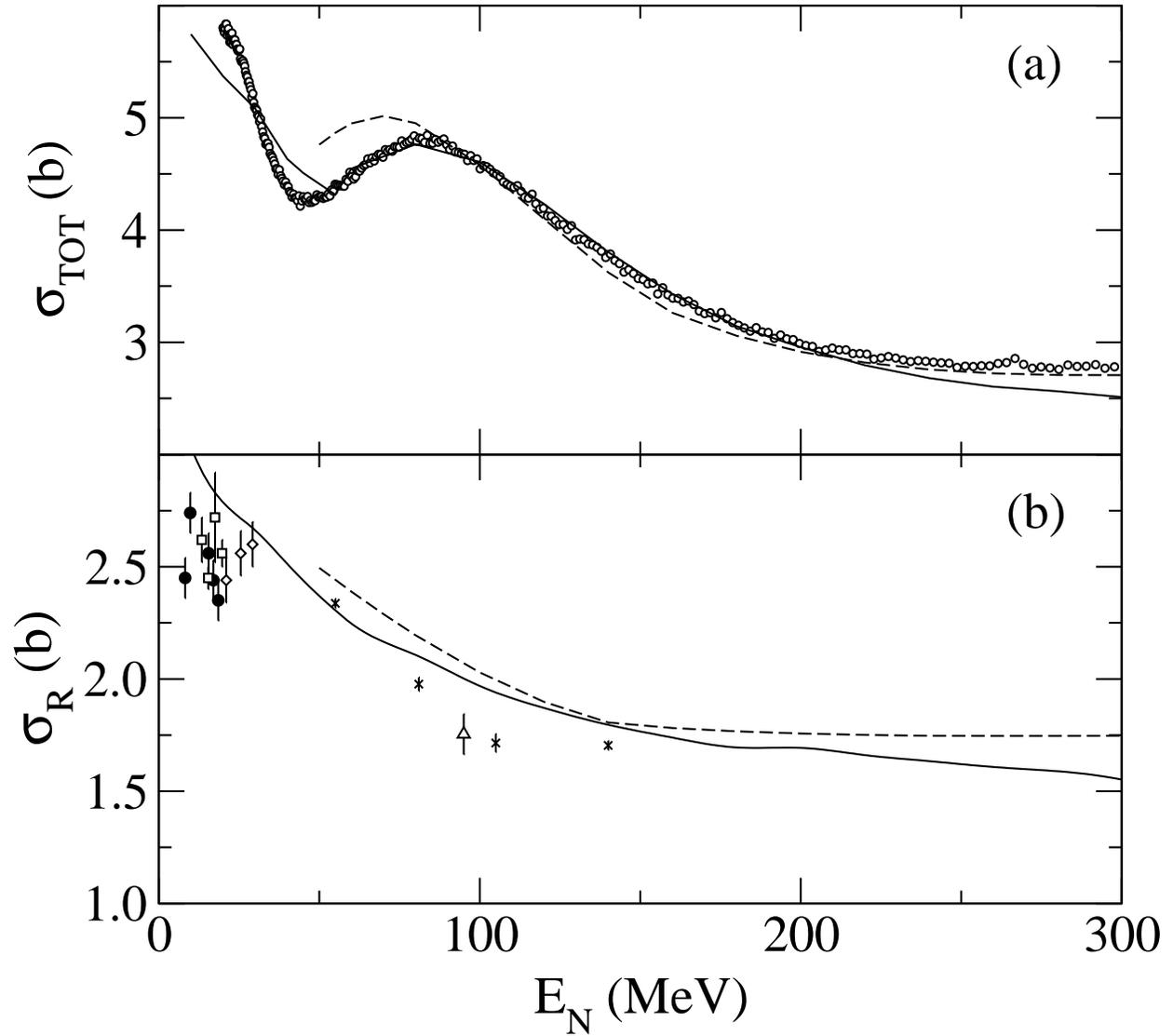}
\caption[]{Total and total reaction cross sections, displayed in (a)
and (b) respectively, for the scattering of neutrons from
$^{208}$Pb. The total cross section data of Findlay {\em et al.} (open
circles) and total reaction cross section data
\cite{Mc58,Vo56,De50,Bo59,De62} are compared to the results of the
microscopic (solid line) and phenomenological (dashed line)
calculations.}
\label{lead-tot}
\end{figure}

In Fig.~\ref{lead}, the results of the model calculations for the
65~MeV scattering of nucleons from $^{208}$Pb are compared to the
proton scattering data of Sakaguchi {\em et al.} (circles) \cite{Sa82}
and the neutron scattering data of Hjort {\em et al.} \cite{Hj94}
(squares) Ibaraki {\em et al.} \cite{Ib00} (diamonds). While both
models give a reasonable representation of the proton scattering data,
with the phenomenological model doing better, the neutron scattering
data are far better reproduced by the microscopic model. The
phenomenological model significantly underpredicts the cross section
above $20^{\circ}$. In Fig.~\ref{lead-tot}, we show the total and
total reaction cross sections for the scattering of neutrons from
$^{208}$Pb. Between 60 and 200~MeV the predicted total cross section
as calculated by the microscopic model shows excellent agreement, to
within 1.5\%, with the data of Finlay {\em et al.}
\cite{De01,Fi93}. The phenomenological model does reasonably well
above 80~MeV, underpredicting the total cross section by at most
4\%. Below 80~MeV, however, the phenomenological model fails to
reproduce the minimum. This is consistent with the relatively poor
results of the phenomenological model at 65~MeV. As the elastic cross
section is well reproduced by the microscopic model, one has
confidence in the predicted microscopic results of the total reaction
cross section. Those measurements rely on careful subtraction of the
elastic from the total cross section and hence some degree of error is
to be expected. That we can predict the reaction cross section to
within 10\% of measurement is encouraging. The phenomenological model
does worse. Note that unlike the optical potentials for the other
nuclei, we find a weaker imaginary part of the microscopic optical
potential relative to that of the phenomenological.

As with the results of Kozack and Madland \cite{Ko90}, which were
based on a relativistic Dirac phenomenological optical potential
model, and of Karataglidis and Madland, as based on the microscopic
Schr\"odinger model \cite{Ka01}, we observe a significant difference
between the proton and neutron analyzing powers for the 65~MeV
scattering from $^{208}$Pb. The analyzing powers are not completely
out of phase, however, as was observed in the other calculations at
100~MeV, suggesting an energy dependence.

\section{Conclusions}

We have presented detailed comparisons of model predictions of proton
and neutron elastic scattering at 65~MeV for a number of nuclei. The
models used were a phenomenological relativistic model and a
microscopic one based on the $g$ matrices of the Bonn-B $NN$ interaction.

In all cases, the differential cross sections and analyzing powers
were well reproduced by both models. This is of note as the
microscopic optical potentials for both proton and neutron scattering
are derived from the same $g$ matrix, with isospin selecting the
appropriate components. Thus the changes in magnitude and shape
between the proton and neutron scattering observables are contained
within the same underlying physics. While the phenomenological
potentials are derived within a relativistic Lane model, those
potentials rely on a global fit to constrain parameters and hence
differences between the proton and neutron potentials. Those
differences are naturally contained in the microscopic model.

Both the microscopic and phenomenological model analyses contained
herein have allowed us to identify a problem with the existing set of
neutron scattering data. The two sets of data (Hjort {\em et al.}
\cite{Hj94} and Ibaraki {\em et al.} \cite{Ib00}) show a discrepancy
between them in the region of the first minimum. The analyses favor
the Ibaraki data and, as the two approaches to the scattering problem
are fundamentally different, one has some confidence in the results. A
new measurement would be required of these data, especially of the
$n$-$^{40}$Ca elastic scattering cross section.

Measurements of the neutron scattering analyzing powers are also
suggested. First, the differences between the two models are most
noticeable in the results for the neutron analyzing powers, suggesting
a method of delineation between the two. Second, for the heavier
nuclei, a significant difference is observed between the proton and
neutron analyzing powers. That difference is attributed to the absence
of the Coulomb potential in neutron scattering \cite{Ko90}. A
measurement of the neutron analyzing power for scattering from
$^{90}$Zr or $^{208}$Pb at intermediate energies, admittedly a
difficult experiment, would demonstrate this effect, and a systematic
study based on the microscopic model is under way to determine in which
cases this effect would be strongest \cite{Ka01a}.

This work was supported by the United States Department of Energy
Contract no. W-7405-ENG-36.

\bibliography{neutron}

\begin{thebibliography}{10}
\providecommand*{\bibinfo}[2]{#2}
\providecommand*{\eprint}[1]{#1}
\providecommand*{\url}[1]{#1}
\bibitem{Review}
\bibinfo{author}{K.~Amos}, \bibinfo{author}{P.~J. Dortmans},
  \bibinfo{author}{H.~V. von Geramb}, \bibinfo{author}{S.~Karataglidis}, and
  \bibinfo{author}{J.~Raynal}, \bibinfo{journal}{Adv. in Nucl. Phys.}
  \bibinfo{volume}{\textbf{25}} (\bibinfo{date}{2000}), 276.
\bibitem{Do98}
\bibinfo{author}{P.~J. Dortmans}, \bibinfo{author}{K.~Amos},
  \bibinfo{author}{S.~Karataglidis}, and \bibinfo{author}{J.~Raynal},
  \bibinfo{journal}{Phys. Rev. C} \bibinfo{volume}{\textbf{58}},
  \bibinfo{pages}{3002} (\bibinfo{date}{1998}).
\bibitem{De01}
\bibinfo{author}{P.~K. Deb}, \bibinfo{author}{K.~Amos},
  \bibinfo{author}{S.~Karataglidis}, \bibinfo{author}{M.~B. Chadwick}, and
  \bibinfo{author}{D.~G. Madland}, \bibinfo{journal}{Phys. Rev. Lett.}
  \bibinfo{volume}{\textbf{86}}, \bibinfo{pages}{3248} (\bibinfo{date}{2001}).
\bibitem{Ko90}
\bibinfo{author}{R.~Kozack} and \bibinfo{author}{D.~G. Madland},
  \bibinfo{journal}{Nucl. Phys.} \bibinfo{volume}{\textbf{A509}},
  \bibinfo{pages}{664} (\bibinfo{date}{1990}).
\bibitem{Ka01}
\bibinfo{author}{S.~Karataglidis} and \bibinfo{author}{D.~G. Madland}
  (\bibinfo{date}{2001}), submitted for publication to Phys. Rev. Lett.,
  nucl-th/0103048.
\bibitem{Ma87}
\bibinfo{author}{R.~Machleidt}, \bibinfo{author}{K.~Holinde}, and
  \bibinfo{author}{C.~Elster}, \bibinfo{journal}{Phys. Rep.}
  \bibinfo{volume}{\textbf{149}}, \bibinfo{pages}{1} (\bibinfo{date}{1987}).
\bibitem{Ra99}
\bibinfo{author}{J.~Raynal}, \bibinfo{title}{\emph{computer code DWBA98}}
  (\bibinfo{date}{1999}), (NEA 1209/05).
\bibitem{Ha70}
\bibinfo{author}{M.~Haftel} and \bibinfo{author}{F.~Tabakin},
  \bibinfo{journal}{Nucl. Phys.} \bibinfo{volume}{\textbf{A158}},
  \bibinfo{pages}{1} (\bibinfo{date}{1970}).
\bibitem{Do91}
\bibinfo{author}{P.~J. Dortmans} and \bibinfo{author}{K.~Amos},
  \bibinfo{journal}{J. Phys. G} \bibinfo{volume}{\textbf{17}},
  \bibinfo{pages}{901} (\bibinfo{date}{1991}).
\bibitem{De00}
\bibinfo{author}{P.~K. Deb} and \bibinfo{author}{K.~Amos},
  \bibinfo{journal}{Phys. Rev. C} \bibinfo{volume}{\textbf{62}},
  \bibinfo{pages}{024605} (\bibinfo{date}{2000}).
\bibitem{Ma97}
\bibinfo{author}{D.~G. Madland}, in \emph{Nucleon-nucleus optical model up to
  200 MeV} (\bibinfo{publisher}{NEA/OECD}, \bibinfo{year}{1997}),
  \bibinfo{pages}{p. 129}.
\bibitem{Sc82}
\bibinfo{author}{P.~Schwandt}, \bibinfo{author}{H.~O. Meyer},
  \bibinfo{author}{W.~W. Jacobs}, \bibinfo{author}{A.~D. Bacher},
  \bibinfo{author}{S.~E. Vigdor}, \bibinfo{author}{M.~D. Kaitchuck}, and
  \bibinfo{author}{T.~R. Donoghue}, \bibinfo{journal}{Phys. Rev. C}
  \bibinfo{volume}{\textbf{26}}, \bibinfo{pages}{55} (\bibinfo{date}{1982}).
\bibitem{Ox86}
OXBASH-MSU (the Oxford-Buenos-Aries-Michigan State University shell model
  code). A. Ecthegoyenm W. D. M. Raye, and N. S. Godwin (MSU version by B. A.
  Brown, 1986); B. A. Brown, A. Etchegoyen, and W. D. M. Rae, MSUCL Report
  Number 524, 1986 (unpublished).
\bibitem{Br00}
\bibinfo{author}{B.~A. Brown}, \bibinfo{journal}{Phys. Rev. Lett.}
  \bibinfo{volume}{\textbf{85}}, \bibinfo{pages}{5296} (\bibinfo{date}{2000}).
\bibitem{Ka85}
\bibinfo{author}{S.~Kato} \emph{et~al.}, \bibinfo{journal}{Phys. Rev. C}
  \bibinfo{volume}{\textbf{31}}, \bibinfo{pages}{1616} (\bibinfo{date}{1985}).
\bibitem{Hj94}
\bibinfo{author}{E.~L. Hjort}, \bibinfo{author}{F.~P. Brady},
  \bibinfo{author}{J.~L. Romero}, \bibinfo{author}{J.~R. Drummond},
  \bibinfo{author}{D.~S. Sorenson}, \bibinfo{author}{J.~H. Osborne},
  \bibinfo{author}{B.~McEachern}, and \bibinfo{author}{L.~F. Hansen},
  \bibinfo{journal}{Phys. Rev. C} \bibinfo{volume}{\textbf{50}},
  \bibinfo{pages}{275} (\bibinfo{date}{1994}).
\bibitem{Ib00}
\bibinfo{author}{M.~Ibaraki}, \bibinfo{author}{M.~Baba},
  \bibinfo{author}{T.~Miura}, \bibinfo{author}{Y.~Hirasawa},
  \bibinfo{author}{Y.~Nauchi}, \bibinfo{author}{H.~Nakashima},
  \bibinfo{author}{S.~Meigo}, \bibinfo{author}{O.~Iwamoto}, and
  \bibinfo{author}{S.~Tanaka}, \bibinfo{journal}{Nucl. Inst. Meth. Phys. Res.}
  \bibinfo{volume}{\textbf{A446}}, \bibinfo{pages}{536} (\bibinfo{date}{2000}),
  M. Baba, private communication.
\bibitem{Sa82}
\bibinfo{author}{H.~Sakaguchi}, \bibinfo{author}{M.~Nakamura},
  \bibinfo{author}{K.~Hatanaka}, \bibinfo{author}{A.~Goto},
  \bibinfo{author}{T.~Noro}, \bibinfo{author}{F.~Ohtani},
  \bibinfo{author}{H.~Sakamoto}, \bibinfo{author}{H.~Ogawa}, and
  \bibinfo{author}{S.~Kobayashi}, \bibinfo{journal}{Phys. Rev. C}
  \bibinfo{volume}{\textbf{26}}, \bibinfo{pages}{944} (\bibinfo{date}{1982}).
\bibitem{Fi93}
\bibinfo{author}{R.~W. Finlay}, \bibinfo{author}{W.~P. Abfalterer},
  \bibinfo{author}{G.~Fink}, \bibinfo{author}{E.~Montei},
  \bibinfo{author}{T.~Adami}, \bibinfo{author}{P.~W. Lisowski},
  \bibinfo{author}{G.~L. Morgan}, and \bibinfo{author}{R.~C. Haight},
  \bibinfo{journal}{Phys. Rev. C} \bibinfo{volume}{\textbf{47}},
  \bibinfo{pages}{237} (\bibinfo{date}{1993}).
\bibitem{Ka01a}
S. Karataglidis and D. G. Madland (to be published).
\bibitem{Vr87}
\bibinfo{author}{H.~de~Vries}, \bibinfo{author}{C.~W. de~Jager}, and
  \bibinfo{author}{C.~de~Vries}, \bibinfo{journal}{At. Data Nucl. Data Tables}
  \bibinfo{volume}{\textbf{36}}, \bibinfo{pages}{495} (\bibinfo{date}{1987}).
\bibitem{Wa92}
\bibinfo{author}{E.~K. Warburton} and \bibinfo{author}{B.~A. Brown},
  \bibinfo{journal}{Phys. Rev. C} \bibinfo{volume}{\textbf{46}},
  \bibinfo{pages}{923} (\bibinfo{date}{1992}).
\bibitem{Br88}
\bibinfo{author}{B.~A. Brown} and \bibinfo{author}{B.~H. Wildenthal},
  \bibinfo{journal}{Ann. Rev. Nucl. Part. Sci.} \bibinfo{volume}{\textbf{36}},
  \bibinfo{pages}{29} (\bibinfo{date}{1988}).
\bibitem{Ri91}
\bibinfo{author}{W.~A. Richter}, \bibinfo{author}{M.~G. van~der Merwe},
  \bibinfo{author}{R.~E. Julies}, and \bibinfo{author}{B.~A. Brown},
  \bibinfo{journal}{Nucl. Phys.} \bibinfo{volume}{\textbf{A523}},
  \bibinfo{pages}{325} (\bibinfo{date}{1991}).
\bibitem{Ji89}
\bibinfo{author}{X.~Ji} and \bibinfo{author}{B.~H. Wildenthal},
  \bibinfo{journal}{Phys. Rev. C} \bibinfo{volume}{\textbf{40}},
  \bibinfo{pages}{389} (\bibinfo{date}{1989}).
\bibitem{Mc58}
\bibinfo{author}{M.~H. MacGregor}, \bibinfo{author}{W.~P. Ball}, and
  \bibinfo{author}{R.~Booth}, \bibinfo{journal}{Phys. Rev.}
  \bibinfo{volume}{\textbf{111}}, \bibinfo{pages}{1155} (\bibinfo{date}{1958}).
\bibitem{Vo56}
\bibinfo{author}{R.~G.~P. Voss} and \bibinfo{author}{R.~Wilson},
  \bibinfo{journal}{Proc. Roy. Soc. A} \bibinfo{volume}{\textbf{236}},
  \bibinfo{pages}{52} (\bibinfo{date}{1956}).
\bibitem{De50}
\bibinfo{author}{J.~Dejuren} and \bibinfo{author}{N.~Knable},
  \bibinfo{journal}{Phys. Rev.} \bibinfo{volume}{\textbf{77}},
  \bibinfo{pages}{606} (\bibinfo{date}{1950}).
\bibitem{Bo59}
\bibinfo{author}{T.~W. Bonner} and \bibinfo{author}{J.~H. Slattery},
  \bibinfo{journal}{Phys. Rev.} \bibinfo{volume}{\textbf{113}},
  \bibinfo{pages}{1088} (\bibinfo{date}{1959}).
\bibitem{De62}
\bibinfo{author}{J.~G. Degtjarev} and \bibinfo{author}{V.~G. Nadtochij},
  \bibinfo{journal}{Sov. J. Nucl. Phys.} \bibinfo{volume}{\textbf{11}},
  \bibinfo{pages}{1043} (\bibinfo{date}{1962}).

\end{thebibliography}

%
%

\begin{table}
\caption[]{Shell model space, interaction, harmonic oscillator
parameter ($b$) and r.m.s. radii for the nuclei listed.}
\label{radii}
\begin{tabular}{cccddc}
Nucleus & Model space & Interaction & $b$ (fm) &
\multicolumn{2}{c}{$r_{rms}$ (fm)} \\
 & & & & (Theory) & (Expt \cite{Vr87}) \\
\hline
$^{12}$C & $(0+2)\hbar\omega$ & WBT \cite{Wa92} & 1.67 & 2.448 &
$2.472 \pm 0.015$
\\
$^{28}$Si & $0\hbar\omega$ & USD \cite{Br88} & 1.85 & 3.088 & $3.086
\pm 0.018$ \\
$^{40}$Ca & $0\hbar\omega$ & packed & 2.00 & 3.464 & $3.482 \pm 0.025$
\\
$^{56}$Fe & $0\hbar\omega^a$ & FPD6 \cite{Ri91} & 2.05 & 3.796 &
$3.801 \pm 0.015$ \\
$^{90}$Zr & NISJ & NISJ \cite{Ji89} & 2.15 & 4.27 & $4.258 \pm 0.008$
\end{tabular}
$^a$ {\footnotesize A 2p-2h $fp$-shell model built on the minimal
$fp$-space wave function.}
\end{table}

\begin{table}
\caption[]{Elastic, total reaction, and total cross sections for the
scattering of 65~MeV neutrons from the nuclei given. The microscopic
and phenomenological results are denoted by MOMP and POMP,
respectively.}
\label{cross}
\begin{tabular}{cddddddc}
Nucleus & \multicolumn{2}{c}{$\sigma_{\text{el}}$ (b)} &
\multicolumn{2}{c}{$\sigma_{\text{R}}$ (b)} &
\multicolumn{3}{c}{$\sigma_{\text{TOT}}$ (b)} \\
 & MOMP & POMP & MOMP & POMP & MOMP & POMP & Expt. \cite{Fi93} \\
\hline
$^{12}$C & 0.417 & 0.537 & 0.331 & 0.269 & 0.748 & 0.806 & $0.753 \pm
0.005$ \\
$^{28}$Si & 0.852 & 0.988 & 0.598 & 0.523 & 1.450 & 1.511 & $1.500 \pm
0.006^a$ \\
$^{40}$Ca & 1.122 & 1.235 & 0.788 & 0.690 & 1.910 & 1.925 & $1.966 \pm
0.009$ \\
$^{56}$Fe & 1.370 & 1.465 & 0.964 & 0.890 & 2.334 & 2.355 & ---    \\
$^{90}$Zr & 1.714 & 1.847 & 1.299 & 1.272 & 3.013 & 3.119 & $3.048 \pm
0.003$ \\
$^{208}$Pb & 2.415 & 2.661 & 2.195 & 2.339 & 4.610 & 5.000 & $4.635
\pm 0.001$
\end{tabular}
$^a$ {\footnotesize A natural target was used.}
\end{table}

%
%

\end{document}